\documentclass[graybox]{svmult}

\usepackage{mathptmx}       
\usepackage{helvet}         
\usepackage{courier}        
\usepackage{type1cm}        
\usepackage{makeidx}         
\usepackage{graphicx}        
\usepackage{multicol}        
\usepackage[bottom]{footmisc}

\usepackage{natbib}
\usepackage{amsmath}
\usepackage{amssymb}
\usepackage{bm}
\usepackage{mathrsfs,dsfont}
\usepackage{rotating}
\usepackage{url}

\newcommand{\indic}{\mathds{1}}

\newcommand{\Y}{\mathbb{Y}}

\newcommand{\R}{\mathds{R}}

\newcommand{\Nbb}{\mathds{N}}
\renewcommand{\P}{\mathds{P}}

\newcommand{\ddr}{\mathrm{d}}
\newcommand{\edr}{\mathrm{e}}

\newcommand{\esp}{\mathds{E}}

\def\simiid{\stackrel{\mbox{\scriptsize{iid}}}{\sim}}

\newcommand{\mt}{\tilde{\mu}} 

\newcommand{\St}{\tilde{S}}
\makeindex             

\begin{document}

\title*{Bayesian Survival Model based on Moment Characterization}
\titlerunning{Bayesian Survival Model}
 \author{Julyan Arbel, Antonio Lijoi and Bernardo Nipoti}
\authorrunning{Arbel et al.}
\institute{Julyan Arbel \at Collegio Carlo Alberto, via Real Collegio, 30, 10024 Moncalieri, Italy, \email{julyan.arbel@carloalberto.org}
\and Antonio Lijoi \at Department of Economics and Management, University of Pavia, Via San Felice 5, 27100 Pavia, Italy, and Collegio Carlo Alberto, via Real Collegio, 30, 10024 Moncalieri, Italy, \email{lijoi@unipv.it}
\and Bernardo Nipoti \at Department of Economics and Statistics, University of Torino, C.so Unione Sovietica 218/bis, 10134 Torino, Italy, and Collegio Carlo Alberto, via Real Collegio, 30, 10024 Moncalieri, Italy, \email{bernardo.nipoti@carloalberto.org}
}
%
%
\maketitle

\abstract{Bayesian nonparametric marginal methods are very popular since they lead to fairly easy implementation due to the formal marginalization of the infinite-dimensional parameter of the model. However, the straightforwardness of these methods also entails some limitations: they typically yield point estimates in the form of posterior expectations, but cannot be used to estimate non-linear functionals of the posterior distribution, such as median, mode or credible intervals. This is particularly relevant in survival analysis where non-linear functionals such as e.g. the median survival time, play a central role for clinicians and practitioners. The main goal of this paper is to summarize the methodology introduced in \citet{arbel2014full} for hazard mixture models in order to draw approximate Bayesian inference on survival functions that is not limited to the posterior mean. In addition, we propose a practical implementation of an \textsf{R} package called \textbf{momentify} designed for moment-based density approximation, and, by means of an extensive simulation study, we thoroughly compare the introduced methodology with standard marginal methods and empirical estimation.
}


\section{Introduction}
\label{Sect:intro}

With \emph{marginal methods} in Bayesian nonparametrics we refer to inferential procedures which rely on the integration (or marginalization) of the infinite-dimensional parameter of the model. This marginalization step is typically achieved by means of the so-called Blackwell--MacQueen P\'olya urn scheme. We consider the popular example of the Dirichlet process \citep[][]{ferguson1973bayesian} to illustrate the idea. Denote by $\bm{Y}=(Y_1,\ldots,Y_n)$ an exchangeable sequence of random variables to which we assign as a prior distribution a Dirichlet process with mass parameter $M$ and base measure $G_0$, that is
\begin{align*}
&Y_i\vert G \simiid G,\\
&G\sim DP(M,G_0).
\end{align*}
The marginal distribution of $\bm{Y}$, once $G$ has been integrated out, can be derived from the set of predictive distributions for $Y_i$, given $(Y_1,\ldots,Y_{i-1})$, for each $i=1,\ldots,n$. In this case, such conditional distributions are linear combinations between the base measure $G_0$ and the empirical distribution of the conditioning variables and are effectively described through a P\'olya urn sampling scheme.
Marginal methods have played a major role in the success of Bayesian nonparametrics since the P\'olya urn generally leads to ready to use Markov chain Monte Carlo (MCMC) sampling strategies which, furthermore, immediately provide Bayesian point estimators in the form of posterior means. A popular example is offered by mixtures of the Dirichlet process for density estimation; 
for the implementation, see e.g. the \textsf{R} package \textbf{DPpackage} by \citet{jara2011dppackage}. 
However, the use of marginal methods has important limitations that we wish to address here. Indeed, one easily notes that the posterior estimates provided by marginal methods are not suitably endowed with measures of uncertainty such as posterior credible intervals. Furthermore, using the posterior mean as an estimator is equivalent to choosing a square loss function which does not allow for other types of estimators such that median or mode of the posterior distribution. Finally, marginal methods do not naturally lead to the estimation of non-linear functionals of the distribution of a survival time, such as the median survival time. For a discussion of these limitations see e.g. \citet{GelKot02}.

The present paper aims at proposing a new procedure that combines closed-form analytical results arising from the application of marginal methods with an approximation of the posterior distribution which makes use of posterior moments. The whole machinery is developed for the estimation of survival functions that are modeled in terms of hazard rate functions. To this end, let $F$ denote the cumulative distribution function (CDF) associated to a probability distribution on $\R^+$. If $F$ is absolutely continuous, the corresponding survival  function and cumulative hazard rate are defined, respectively, by $S(t)=1-F(t)$ and $H(t)=-\log (S(t))$. Then, the hazard rate function is given by $h(t)=-S'(t)/S(t)$. Let us recall that survival analysis has been a very active  area of application of Bayesian nonparametric methodology: neutral to the right processes were used by \cite{Doksum} as a prior for the CDF $F$,  and beta processes by \cite{Cervo} as a prior for the cumulative hazard function $H$, both benefiting from useful conjugacy properties. Here we specify a prior on the hazard rate $h$. The most popular example is the gamma process mixture, originally proposed in \cite{DykLau81}. More general models have been studied in later work by \cite{LoWen89} and \cite{Jam05}. 
Bayesian inference for these models often relies on a marginal method, see, e.g., \cite{IshJam04}. Though quite simple to implement, marginal methods typically yield estimates of the hazard rate, or equivalently of the survival function, only in the form of posterior mean at a fixed time point. Working along the lines of \citet{arbel2014full}, we show that a clever use of a moment-based approximation method does provide a relevant upgrade on the type of inference one can draw via marginal sampling schemes. We should stress that the information gathered by marginal methods is not confined to the posterior mean but is actually much richer and, if properly exploited, can lead to a more complete posterior inference. 

Let us briefly introduce Bayesian hazard mixture models. Random parameters, such as the hazard rate and survival function, are denoted with a tilde on top, e.g. $\tilde h$ and $\tilde S$. We endow $\tilde h$ with a prior distribution defined by the distribution of the random hazard rate (RHR)
\begin{equation}\label{RHR}
\tilde h(t)=\int_\Y k(t;y)\mt(\ddr y),
\end{equation}
where $\mt$ is a completely random measure (CRM) on $\Y=\R^+$, and $k(\,\cdot\,;\cdot\,)$ denotes a transition kernel on $\R^+\times \Y$. Under suitable assumption on the CRM $\tilde \mu$, we have $\lim_{t\rightarrow \infty} \int_0^t \tilde h(s)\ddr s=\infty$ with probability 1. Therefore we can adopt the following model 
\begin{equation}
\label{eq:exch_model}
  \begin{split}
    X_i\,|\,\tilde P \: &\simiid \: \tilde P\\
    \tilde P((t,\infty)) \:    &\stackrel{\scriptsize{\mbox{d}}}{=}\: \tilde S(t) \:\stackrel{\scriptsize{\mbox{d}}}{=}\: \exp\left(-\int_0^{t}
      \tilde h(s)\,\ddr s\right)
  \end{split}
\end{equation}
for a sequence of (possibly censored) survival data $\bm{X} = (X_1,\ldots,X_n)$. 
In this setting, \cite{DykLau81} characterize the posterior distribution of the so-called \textit{extended gamma process}: this is obtained when $\mt$ is a gamma CRM and $k(t;y)=\indic_{(0,t]}(y)\,\beta(y)$ for some positive right-continuous function $\beta :\R^+\to\R^+$. The same kind of result is proved in \cite{LoWen89} for \textit{weighted gamma processes} corresponding to RHRs obtained when $\mt$ is still a  gamma CRM and $k(\,\cdot\,;\,\cdot\,)$ is an arbitrary kernel. Finally, a posterior characterization has been derived by \cite{Jam05} for any CRM $\mt$ and kernel $k(\,\cdot\,;\,\cdot\,)$. 

The rest of the paper is organized as follows. In Section~\ref{sec:mom}, we provide 
the closed-form expressions for the posterior moments of the survival function. We then show in Section~\ref{sec:approx} how to exploit the expression for the moments to approximate the corresponding density function and sample from it. Finally, in Section~\ref{sec:inference} we study the performance of our methodology by means of an extensive simulation study with survival data.

\section{Moments of the posterior survival function}\label{sec:mom}


Closed-form expressions for the moments of any order of the posterior survival curve $\tilde S(t)$ at any $t$ are provided in \citet{arbel2014full}. For a complete account, we recall the result hereafter. We first need to introduce some notation. A useful augmentation suggests introducing latent random variables $\bm{Y}=(Y_1,\ldots,Y_n)$ such that, building upon the  posterior characterization derived by \cite{Jam05}, we can derive expressions for the posterior moments of the random variable $\St(t)$, where $t$ is fixed, conditionally on $\bm{X}$ and $\bm{Y}$. To this end, define  $K_x(y) = \int_0^{x}k(s;y)\ddr s$ and $K_{\bm{X}}(y)=\sum_{i=1}^n K_{X_i}(y)$. Also, the almost sure discreteness of $\tilde \mu$ implies there might be ties among the $Y_i$'s with positive probability.
Therefore, we denote the distinct values among $\bm{Y}$ by $(Y_1^*,\ldots,Y_k^*)$, where $k\leq n$,  and, for any $j=1,\ldots,k$, we define $C_j=\left\{l\,:\,Y_l=Y_j^*\right\}$ and $n_j=\#\,C_j$. We can now state the following result.
\begin{proposition}\label{mom} Denote by $\nu(\ddr s,\ddr y) = \rho(s)\,\ddr s\,c\,P_0(\ddr y)$ the \textit{L\'evy intensity} of the completely random measure $\mt$. Then for every $t>0$ and $r>0$,
\begin{multline}
\esp[\St^r(t)\,|\,\bm{X},\bm{Y}]=\exp\left\{-\int_{\R^+\times \Y}\left(1-\edr^{-r K_t(y) s}\right)\edr^{-K_{\bm{X}}(y) s} \nu(\ddr s,\ddr y) \right\}\\
\times\prod_{j=1}^k \frac{1}{\mbox{B}_j}\int_{\R^+}\exp\left\{-s\left(r K_t(Y_j^*)+K_{\bm{X}}(Y_j^*)\right)\right\}s^{n_j}\rho(s)\ddr s,
\end{multline}
where $B_j=\int_{\R^+} s^{n_j} \exp\left\{-s K_{\bm{X}}(Y_j^*)\right\}\rho(s) \ddr s$, for $j=1,\ldots,k$.
\end{proposition} 

For evaluating the posterior moments $\esp[\St^r(t)\,|\,\bm{X}]$ by means of Proposition~\ref{mom}, we use a Gibbs sampler which proceeds by alternately sampling, at each iteration $\ell=1,\ldots,L$, from the full conditional distributions of the latent variables $\bm{Y}$ and the parameters of the model, 
and evaluate $\esp[\St^r(t)\,|\,\bm{X},\bm{Y}]^{(\ell)}$ at each step. For an exhaustive description of the posterior sampling and the expression of the full conditional distributions, see \citet{arbel2014full}.  The remaining of the paper is devoted to illustrating how the characterization of the moments provided by Proposition~\ref{mom} can be used to approximate a density function and, in turn, to carry out Bayesian inference. 

\section{Moment-based density approximation}\label{sec:approx}

The aim is to recover the posterior distribution of the random variable $\tilde S(t)$ for any fixed $t$, based on the knowledge of its moments $\esp[\St^r(t)\,|\,\bm{X}]$ obtained from Proposition~\ref{mom}. In order to simplify the notation, let us consider a generic continuous random variable $S$ on $[0,1]$, and denote by $f$ its density, and its raw moments by $\gamma_r=\esp\big[S^r\big]$, with $r\in\Nbb$. Recovering $f$ from the explicit knowledge of its moments $\gamma_r$
is a classical problem in probability and statistics that has received great attention in the literature. 
See e.g. \cite{provost2005moment} and references and motivating applications therein. A very general approach relies on the basis of Jacobi polynomials $(G_i(s) = \sum_{r=0}^i G_{i,r}s^r)_{i\geq 1}$. They constitute a broad class which includes, among others, Legendre and Chebyshev polynomials, and which is well suited for the expansion of densities with compact support \citep[see][]{provost2005moment}. Any univariate density $f$ supported on $[0,1]$ can be uniquely decomposed on such a basis and therefore there is a unique sequence of real numbers $(\lambda_i)_{i \geq 0}$ such that $f(s)=w_{a,b}(s)\sum_{i=0}^\infty \lambda_i G_i(s)$ where $w_{a,b}(s)=s^{a-1}(1-s)^{b-1}$ 
is named the \emph{weight function} of the basis and is proportional to a beta density in the case of Jacobi polynomials. From the evaluation of  
$\int_0^1 f(s)\, G_i(s)\,\ddr s$ it follows that each $\lambda_i$ coincides with a linear combination of the first $i$ moments of $S$, specifically $\lambda_i=\sum_{r=0}^i G_{i,r}\gamma_r$. 
Then, the polynomial approximation method consists in truncating the representation of $f$ in the Jacobi basis at a given level $i=N$. This procedure leads to a methodology that makes use only of the first $N$ moments and provides the approximation
\begin{equation}\label{eq:approx}
f_N(s)=w_{a,b}(s)\sum_{i=0}^N \left(\sum_{r=0}^i G_{i,r}\mu_r\right) G_i(s).
\end{equation}
It is important to stress that the polynomial  approximation~\eqref{eq:approx} is not necessarily a density as it might fail to be positive or to integrate to 1. In order to overcome this problem, we consider the density $\pi$ proportional to the positive part of~\eqref{eq:approx} defined by $\pi(s)\propto\max(f_N(s),0)$. We resort to the \emph{rejection sampler} 
for sampling from $\pi$. 
This is a method for drawing independently from a distribution proportional to a given non-negative function, that exempts us from computing the normalizing constant corresponding to $\pi$. More precisely, the method requires to pick a proposal distribution $p$ for which there exists a positive constant $M$ such that $\pi\leq M p$. A natural choice for $p$ is the beta distribution proportional to the weight function $w_{a,b}$. Approximation~\eqref{eq:approx} and the rejection sampler were implemented in \textsf{R}. For the purpose of the paper, we have wrapped up the corresponding code in an \textsf{R} package called \textbf{momentify}.\footnote{The \textbf{momentify} package can be downloaded from the first author's (JA) webpage\\ \url{http://www.crest.fr/pagesperso.php?user=3130}.}
In Sections~\ref{pack} and \ref{simul} we briefly describe the package implementation and give a simple working example.

\subsection{Package implementation}\label{pack}

The  major function in package \textbf{momentify}  is called \verb!momentify! and allows for (i) approximating a density based on its moments, and (ii) sampling from this approximate distribution by using the rejection sampler.
The synopsis of the function, together with default values, are given by
\begin{center}
\verb!momentify(moments, N_moments = length(moments), !\\
     \verb! N_sim = 1000, xgrid = seq(0, 1, length = 200))!
\end{center}
The only required argument is \verb!moments!, a $d$-dimensional vector, with $d\geq 2$, composed by the values of the first $d$ consecutive raw moments. The remaining arguments are optional: \verb!N_moments! corresponds to $N$, the number of moments to be used (where $N\leq d$), \verb!N_sim! is the size of the sample obtained by the rejection sampler, and \verb!xgrid! denotes the grid on which the density is to be approximated. 

The function returns a list, say  \verb!res!, with the following components: \verb!xgrid!, defined in argument, \verb!approx_density!, the approximated density evaluated on \verb!xgrid!, and \verb!psample!, the sample obtained from \verb!approx_density! by the reject algorithm. The class of the output list \verb!res! is called \verb!momentify!. For visualizing the output  \verb!res!, two method functions can be readily applied to this class, namely
\verb!plot(res, ...)! and
\verb!hist(res, ...)!.

\subsection{Simulated example}\label{simul}

\begin{center}
\begin{figure}[ht!]
\includegraphics[width=\linewidth]{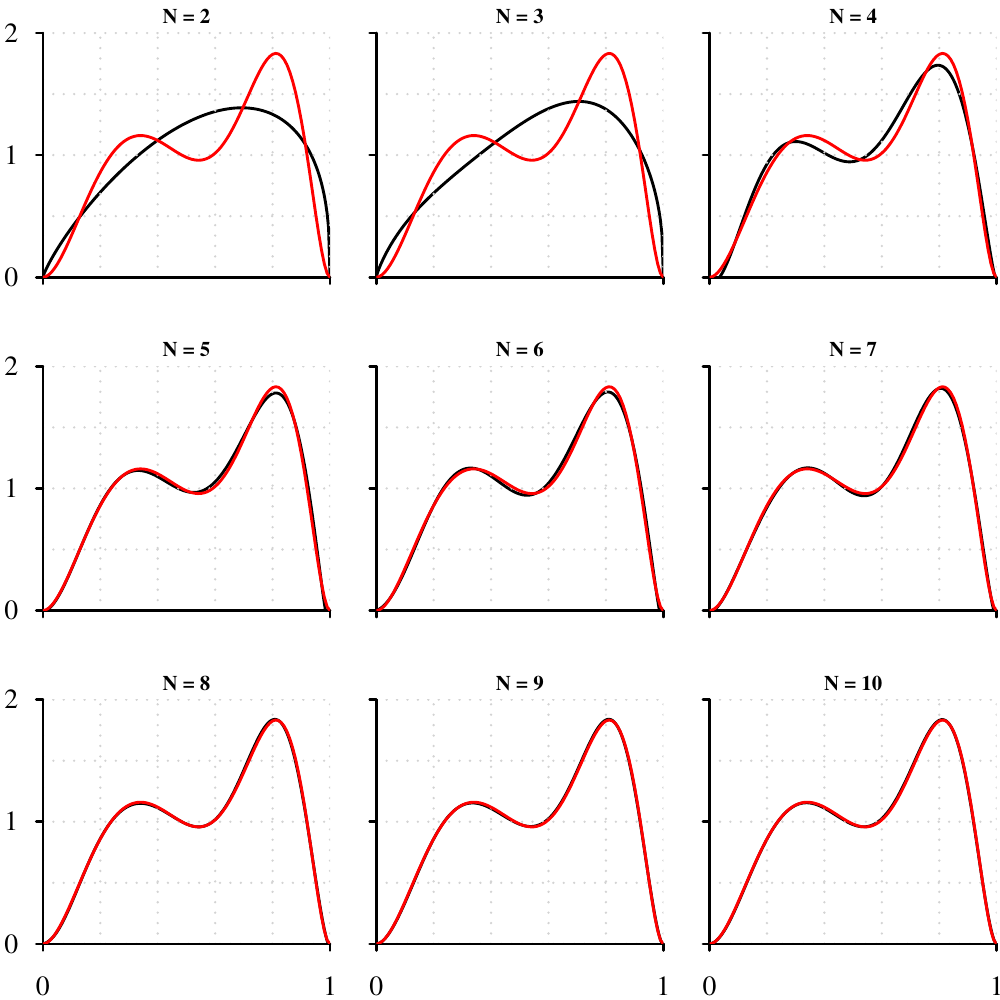}
\caption{Output of \textbf{momentify} \textsf{R} package. True density $f$ of $S$ (in red) and approximated density $f_N$ (in black) involving an increasing number of moments, from $N=2$ (top left) to $N=10$ (bottom right).}
\label{fig:momentify}
\end{figure}
\end{center}

We assess  now the quality of this approximation procedure on a particular example by means of a practical implementation of the \textbf{momentify} package. We specify the distribution of the random variable $S$ by a mixture, with weights of 1/2, of beta distributions of parameters $(a,b)=(3,5)$ and $(c,d)=(10,3)$. The raw moments of any order of $S$ can be explicitly evaluated by
\begin{equation*}
\gamma_r = \esp[S^r] = \frac{a_{(r)}}{(a+b)_{(r)}}  + \frac{c_{(r)}}{(c+d)_{(r)}}
\end{equation*}
where $x_{(r)}=\Gamma(x+r)/\Gamma(x)$. 
As describe above, given a vector of $N$ moments $(\gamma_1,\ldots,\gamma_N)$, the introduced package allows us to approximately evaluate the density~\eqref{eq:approx} and, in turn, to compare it with the true density. The corresponding code for $N=2,\ldots,10$ is the following:
\begin{verbatim}
rfun=function(n){bin=rbinom(n,1,.5)
     bin*rbeta(n,3,5)+(1-bin)*rbeta(n,10,3)}
true_density=function(n){.5*dbeta(n,3,5)+
                     .5*dbeta(n,10,3)}
sim_data = rfun(10^5)
moments = mean(sim_data)
for (i in 2:10){
  moments = c(moments,mean(sim_data^i))
  res = momentify(moments = moments)
  plot(res, main = paste("N =",i))
  curve(true_density(x),add=TRUE, col = "red")
}
\end{verbatim}
The graphical output is given in Figure~\ref{fig:momentify}. We can see that four moments are needed in order to capture the two modes of the distribution, although coarsely. From seven  moments onward, the fit is very good since the two curves are hardly distinguishable. Following this example as well as other investigations not reported here, we choose, as a rule of thumb, to work with $N=10$ moments. A more elaborated numerical study is presented in \citet{arbel2014full} in the context of survival analysis.
%
\section{Bayesian inference}\label{sec:inference}

\subsection{Estimation of functionals of $\St$}\label{sec:functionals}

Given a sample of survival times $\bm{X}=\{X_1,\ldots,X_n\}$, we estimate the first $N$ moments of the posterior distribution of $\tilde S(t)$, for $t$ on a grid of $q$ equally-spaced points $\{t_1,\ldots,t_{q}\}$ in an interval $[0,M]$ by using the Gibbs sampler succinctly described in Section~\ref{sec:mom}. We then exploit the estimated moments to sample from an approximation of the posterior distribution of $\tilde S(t_i)$ for $i=1,\ldots,q$ according to the methodology set forth in Section~\ref{sec:approx}. This allows us to carry out Bayesian inference, with a focus on the estimation of the median survival time and, for any given $t$ in the grid, of credible intervals for $\tilde S(t)$.
The same approach can be easily used to estimate the posterior median and mode of $\tilde S(t)$ at any given $t$, and, in line of principle, any functional of interest.  Let us first consider the median survival time that we denote by $m$. The identity for the cumulative distribution function of $m$, $\P\left(m\leq t\vert\bm{X}\right) = \P\big(\tilde S(t) \leq 1/2 \vert\bm{X}\big)$, allows us to evaluate the CDF of $m$ at each time point $t_i$ as $c_i=\P\big(\tilde S(t_i) \leq 1/2 \vert\bm{X}\big)$. Then, we can estimate the median survival time $m$ by means of the following approximation:
\begin{equation}\label{mst}
\hat m=\esp_{\bm{X}}[m]=\int_0^\infty \P[m>t\vert\bm{X}]\:\ddr t\approx \frac{M}{q-1}\sum_{i=1}^q(1-c_i)
\end{equation}
where the subscript $\bm{X}$ in $\esp_{\bm{X}}[m]$ indicates that the integral is with respect to the distribution of $\tilde S(\cdot)$ conditional to $\bm{X}$. Moreover, the sequence $(c_i)_{i=1}^q$ can be used to devise credible intervals for the median survival time. Similarly, the posterior samples generated by the rejection sampler can be easily used to devise,  $t$-by-$t$, credible intervals for $\tilde S(t)$ or to estimate other functionals that convey meaningful information such as the posterior mode and median. In Section~\ref{sec:simulated}, we apply this methodology in a study involving simulated survival data where we compare the performance of the moment-based methodology with standard marginal methods.
\subsection{Applications}\label{sec:simulated}
For the purpose of illustration, we complete the model specification by assuming a Dykstra and Laud type of kernel $k(t;y)=\indic_{(0,t]}(y)\beta$, for some constant $\beta>0$, a gamma CRM $\tilde\mu$ and an exponential base measure $P_0$ with rate parameter $3$. Moreover, for the hyperparameters $c$ and $\beta$ we choose independent gamma prior distributions with shape parameter $1$ and rate parameter $1/3$. We then consider three samples $\bm{X} = (X_1,\ldots,X_n)$ of size $n=20,100,500$ from a Weibull distribution of parameters $(2,2)$ whose survival function is  $S_0(t)=\exp(-t^2/4)$.
We set $M=6$ (the largest observation in the samples is 5.20) and $q=50$ for the analysis of each sample. We approximately evaluate, $t$-by-$t$, the posterior distribution of $\tilde S(t)$ together with the posterior distribution of the median survival time $m$. 
\begin{figure}
\centering
\includegraphics[width=.328\linewidth]{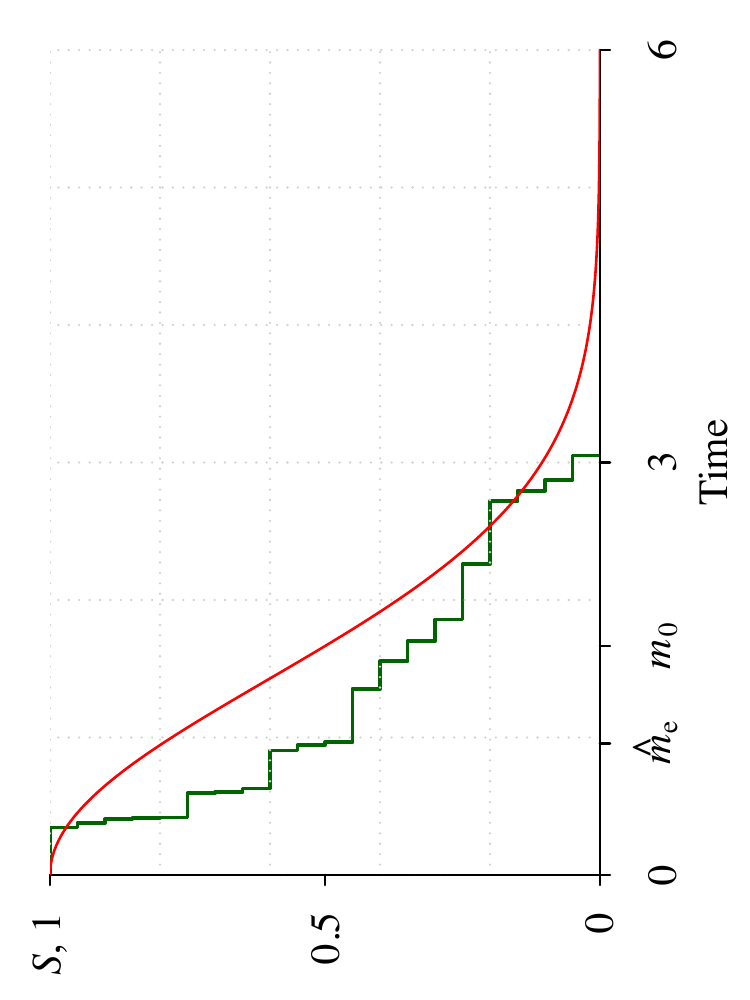}
\includegraphics[width=.328\linewidth]{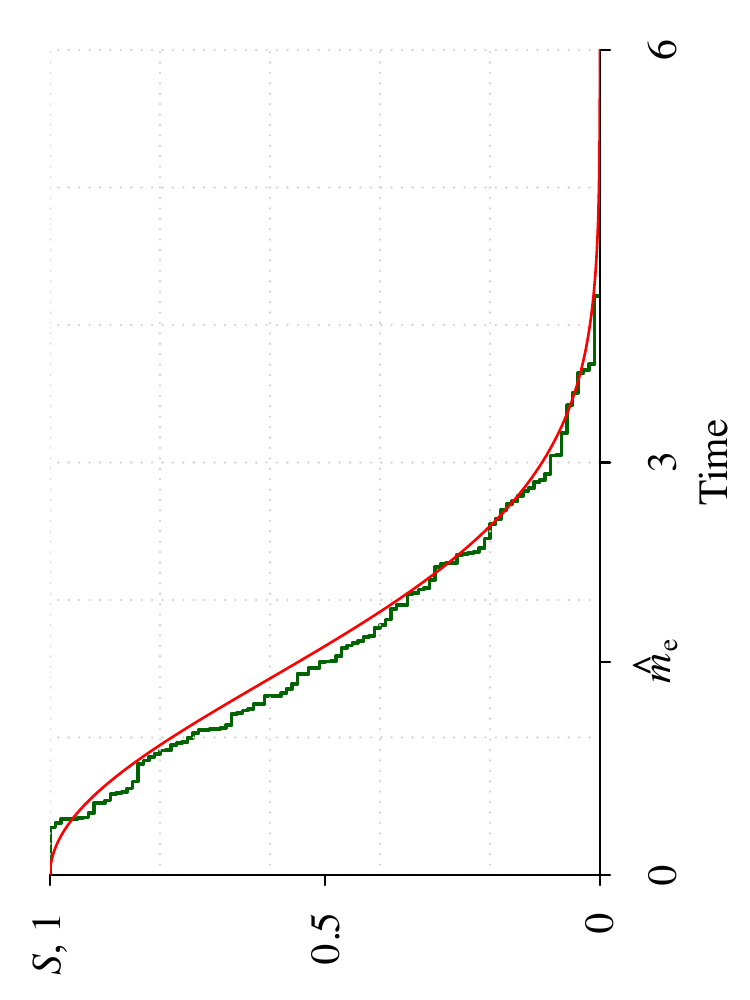}
\includegraphics[width=.328\linewidth]{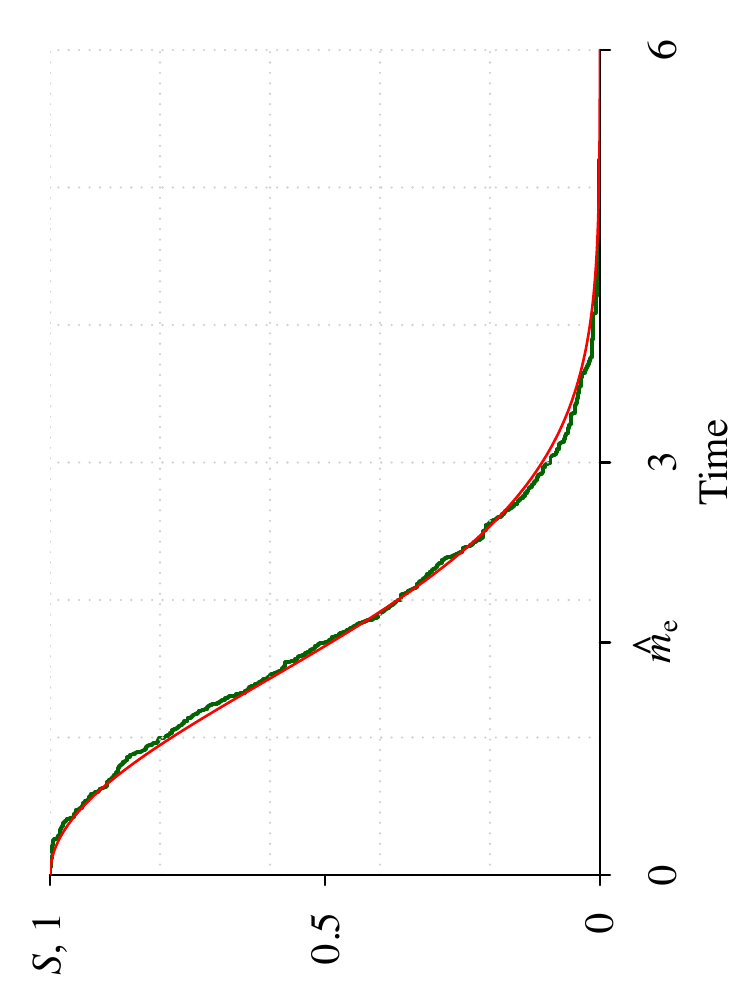}
\includegraphics[width=.328\linewidth]{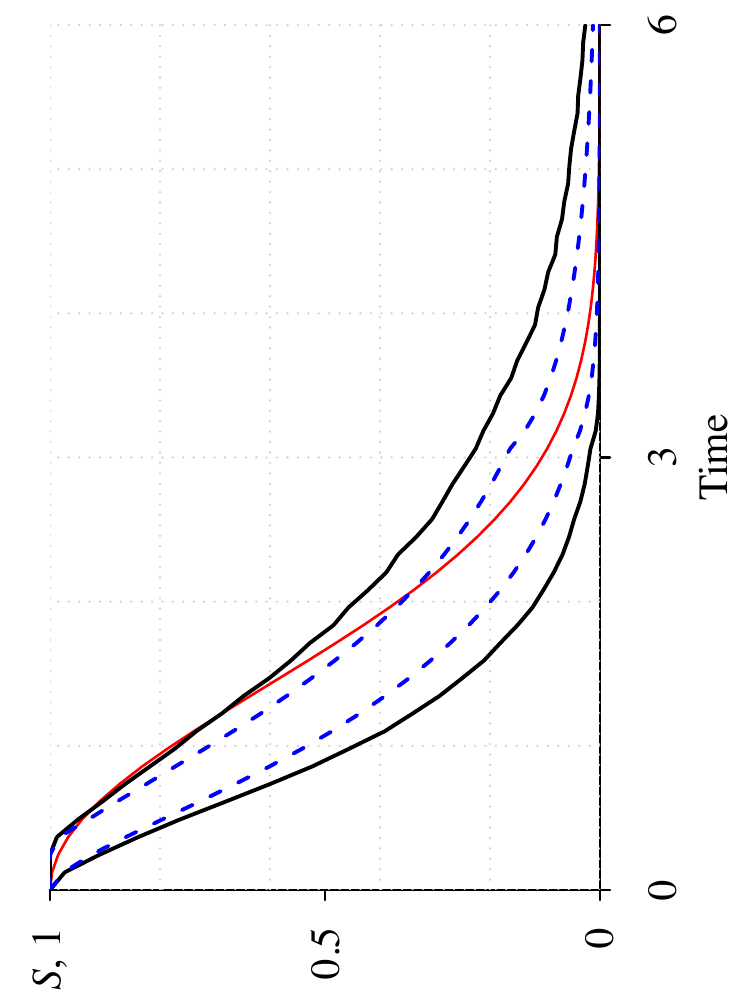}
\includegraphics[width=.328\linewidth]{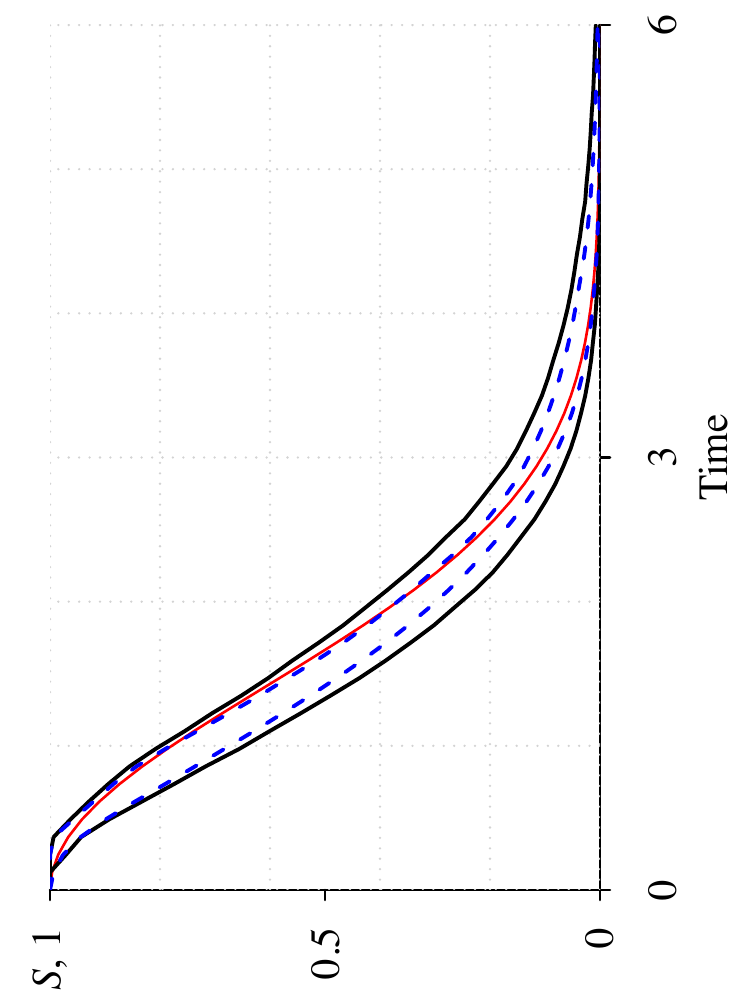}
\includegraphics[width=.328\linewidth]{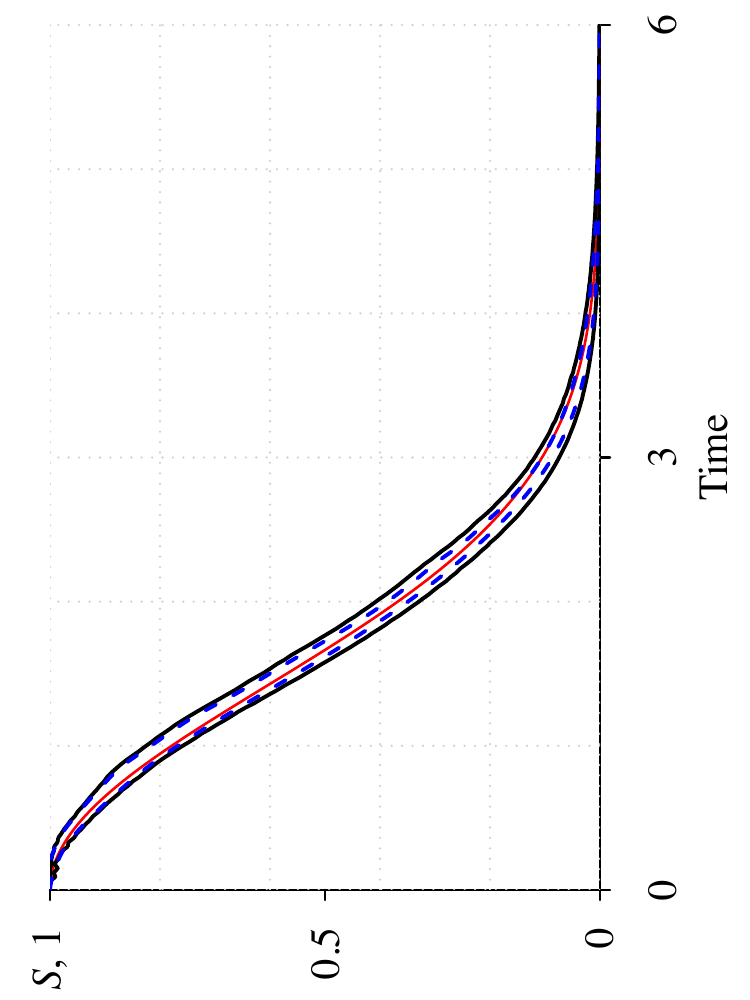}
\includegraphics[width=.328\linewidth]{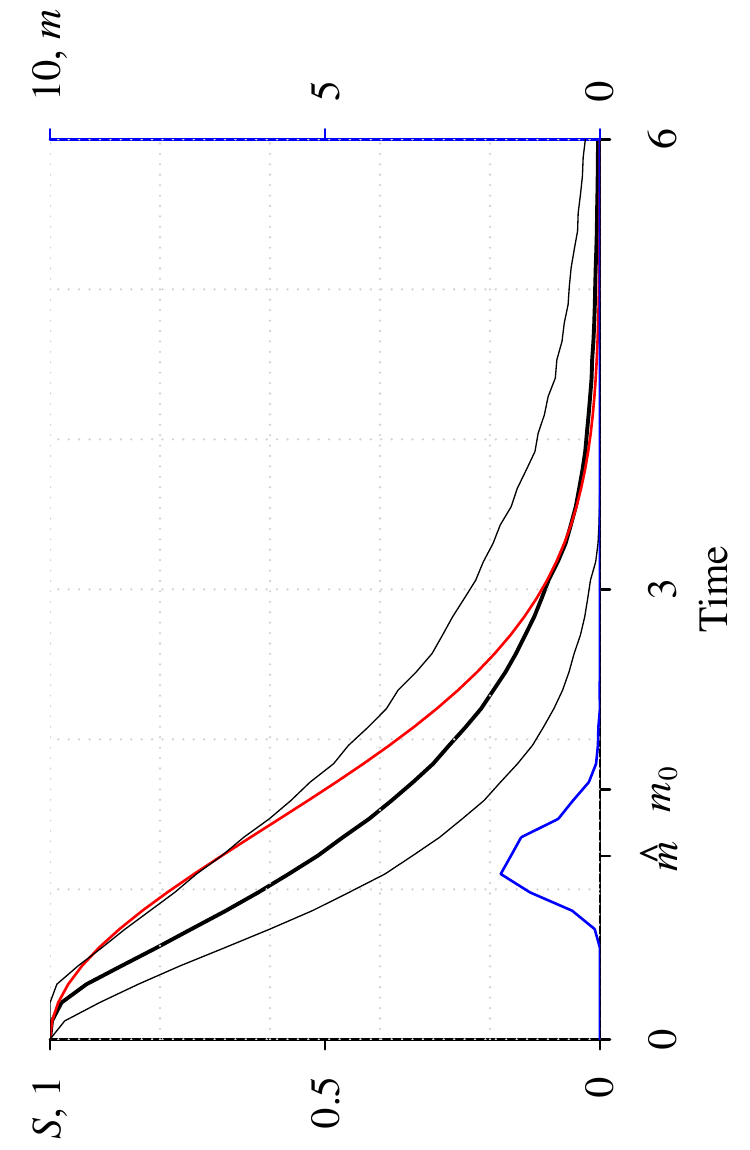}
\includegraphics[width=.328\linewidth]{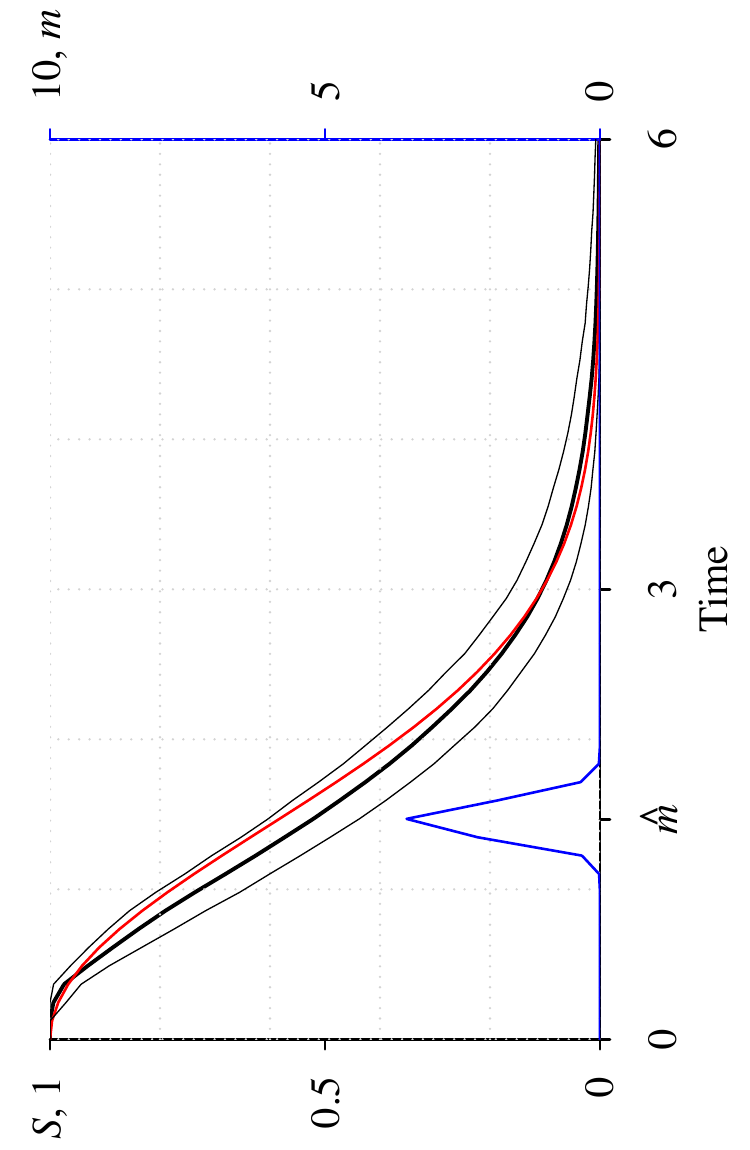}
\includegraphics[width=.328\linewidth]{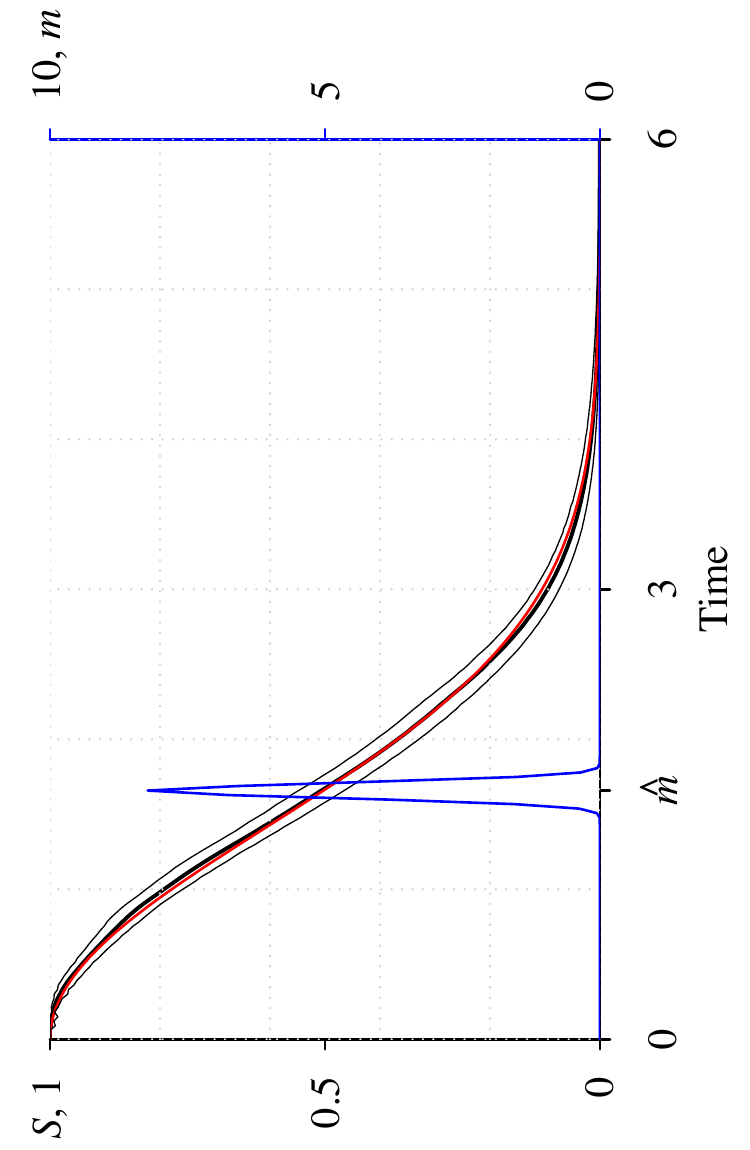}
\caption{The true survival function $S_0(t)$ is the red line in all plots. Bottom row: estimated posterior mean (black solid line) with 95\% credible intervals for $\tilde S(t)$ (black thin lines); in blue the posterior distribution of the median survival time $m$. Middle row: comparison of the 95\% credible interval (black line) with the marginal interval (dashed blue line). Top row: Kaplan--Meier estimate (green line).    
Sample size $n=20$ (left column), $n=100$ (middle column), $n=500$ (right column).}
\label{fig:simulated_examples}
\end{figure}
By inspecting the bottom row of Figure~\ref{fig:simulated_examples}, we can appreciate that the estimated credible intervals for $\tilde S(t)$ contain the true survival function. Moreover, the posterior distribution of the median survival time $m$ (blue curve) is nicely concentrated around the true value $m_0$. 
When relying on marginal methods, the most natural choice for quantifying the uncertainty of posterior estimates consists in considering the quantile  intervals corresponding to the output of the Gibbs sampler, that we refer to as \emph{marginal  intervals}. This leads to consider, for any fixed $t$, the interval whose lower and upper extremes are the quantiles of order e.g. $0.025$ and $0.975$, respectively, of the sample of posterior means
$\big(\esp[\St(t)\,|\,\bm{X},\bm{Y}]^{(\ell)}\big)_{\ell=1,\ldots,L}$ obtained, conditional on $\bm{Y}$, by the Gibbs sampler described in Section~\ref{sec:mom}. In the middle row of Figure~\ref{fig:simulated_examples} we have compared the estimated 
95\% credible intervals 
for $\tilde S(t)$ (black)
and the marginal  intervals corresponding to the output of the Gibbs sampler (dashed blue). In this example, the credible intervals in general contain the true survival function $S_0(t)$, while this does not hold for the marginal intervals. This fact suggests that the marginal method tends to underestimate the uncertainty associated to the posterior estimates, and can be explained by observing that, since the underlying CRM is marginalized out, the intervals arising from the Gibbs sampler output capture only the variability of the posterior mean that can be traced back to the 
latent variables $\bm{Y}$ and the parameters $(c,\beta)$. As a result, especially for a small sample size, the uncertainty detected by the marginal method leads to marginal intervals that can be significantly narrower than the actual posterior credible intervals that we approximate through the moment-based approach. 
The Kaplan--Meier estimates of $\St(t)$ are plotted on the top row of Figure~\ref{fig:simulated_examples}.
\begin{figure}[ht!]
\begin{center}
\includegraphics[width=.9\linewidth]{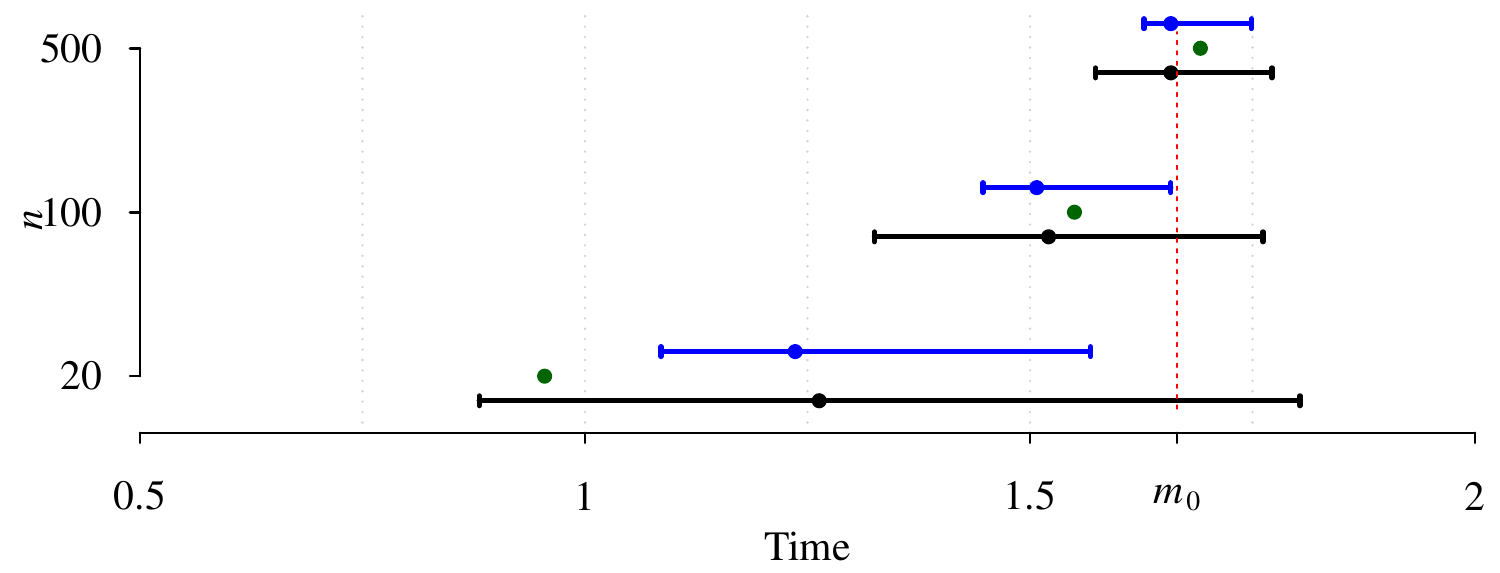}
\caption{Comparison of credible intervals for the median survival time $m$ obtained with the moment-based approach (black line, below for each $n$) and  marginal intervals (blue line, above for each $n$), for varying sample size $n$. The dots indicate the estimators ($\hat m$ in black, $\hat m_m$ in blue and $\hat m_e$ in green). The true median $m_0 = 2\sqrt{\log 2}\approx 1.665$ is indicated by the vertical red dashed line.}
\label{fig:medians}
\end{center}
\end{figure}
\begin{table}[h!]
\caption{Comparison of the median survival time estimated by means of the moment-based method, $\hat m$, by means of the marginal method, $\hat m_m$, and the empirical median survival time $\hat m_e$, for different sample sizes $n$. For the moment-based estimation we show $\hat m$, the absolute error $|\hat m-m_0|$ and $95\%$ credible interval ($CI$); for the marginal method we show $\hat m_m$, the absolute error $|\hat m_m-m_0|$ and the 95\% marginal interval ($CI_m$); last two columns show the empirical estimate $\hat m_e$ and the corresponding absolute error $|\hat m_e - m_0|$. The true median survival time  is $m_0 = 2\sqrt{\log 2}\approx 1.665$.}
\begin{center}
\begin{tabular}{|c|ccccc|ccccc|cc|}
  \hline
\multicolumn{1}{|c}{}  &\multicolumn{5}{c}{moment-based method} &\multicolumn{5}{c}{Marginal} & \multicolumn{2}{c|}{Empirical}\\
\hline
$n$ & $\hat m$  & $\vert\hat m-m_0\vert$ & \multicolumn{3}{c|}{$CI$} & $\hat m_m$  & $\vert\hat m_m-m_0\vert$ & \multicolumn{3}{c|}{$CI_m$} & $\hat m_e$ & $\vert\hat m_e-m_0\vert$ \\
  \hline
  20 & 1.26 & 0.40 & 0.88 & - & 1.80 & 1.24 & 0.43 & 1.09 & - & 1.57 & 0.95 & 0.71 \\ 
  100 & 1.52 & 0.14 & 1.33 & - & 1.76 & 1.51 & 0.16 & 1.45 & - & 1.66 & 1.55 & 0.12 \\ 
  500 & 1.66 & 0.01 & 1.57 & - & 1.77 & 1.66 & 0.01 & 1.63 & - & 1.75 & 1.69 & 0.03 \\ 
   \hline
\end{tabular}\label{table_mst}
   \end{center}
   \end{table}

As described in Section~\ref{sec:functionals}, the moment-based approach enables us to approximate the posterior distribution of the median survival time $m$ (in blue in the bottom row of Figure~\ref{fig:simulated_examples}). This, in turn, can be used to derive sensible credible intervals for $m$. On the other side, when relying on marginal methods, the posterior of the median survival time is not available \textit{per se}. However, in the same way as we defined \textit{marginal intervals} in place of credible intervals for the survival function $\tilde S(t)$, for every $t_i$ the Gibbs sample $\big(\esp[\St(t_i)\,|\,\bm{X},\bm{Y}]^{(\ell)}\big)_{\ell=1,\ldots,L}$ can be used as a proxy of a posterior sample for $\St(t_i)$ in order to provide the following approximation of the CDF of $m$:
\begin{equation}\label{eq:marginal-for-m}
\P\left(m\leq t\vert\bm{X}\right) \approx 
\frac{1}{L}\#\{\ell\,:\,\esp[\St(t)\,|\,\bm{X},\bm{Y}]^{(\ell)}\leq 1/2\}.
\end{equation}
As in \eqref{mst}, an estimator for the median survival time can be obtained as the mean of the distribution whose CDF is given in \eqref{eq:marginal-for-m}. We call such estimator $\hat m_m$ to denote the fact that it is obtained by means of a marginal method. Similarly, from~\eqref{eq:marginal-for-m}, marginal intervals for $\hat m_m$ can be derived as described in Section~\ref{sec:functionals}.   
Finally, we denote by $\hat m_e$ the empirical estimator of $m$ and by $m_0 = 2\sqrt{\log 2}\approx 1.665$ the true median survival time. We summarize the estimates we obtained for the median survival time $m$ in Figure~\ref{fig:medians} and in Table~\ref{table_mst}. For all the sample sizes considered, the credible intervals for $\hat m$ contain the true value. Moreover, as expected, when $n$ grows they shrink around $m_0$: for instance the length of the interval reduces from 0.92 to 0.20 when the sample size $n$ increases from 20 to 500. As observed for the marginal intervals $\tilde S(t)$ at a given $t$, the marginal intervals for $\hat m_m$ obtained with the marginal method and described in Equation~\eqref{eq:marginal-for-m} are in general narrower than the credible intervals obtained by the moment-based approach. Moreover, in this example, they contain the true $m_0$ only for $n=500$. This observation suggests that the use of intervals produced by marginal methods as proxies for posterior credible intervals should be avoided, especially for small sample sizes.


\subsubsection*{Acknowledgment}
J. Arbel and A. Lijoi are supported by the European Research Council (ERC) through StG ``N-BNP'' 306406.

\bibliographystyle{spbasic}
\bibliography{hazbib}

\begin{thebibliography}{11}
\providecommand{\natexlab}[1]{#1}
\providecommand{\url}[1]{{#1}}
\providecommand{\urlprefix}{URL }
\expandafter\ifx\csname urlstyle\endcsname\relax
  \providecommand{\doi}[1]{DOI~\discretionary{}{}{}#1}\else
  \providecommand{\doi}{DOI~\discretionary{}{}{}\begingroup
  \urlstyle{rm}\Url}\fi
\providecommand{\eprint}[2][]{\url{#2}}

\bibitem[{Arbel et~al(2015)Arbel, Lijoi, and Nipoti}]{arbel2014full}
Arbel J, Lijoi A, Nipoti B (2015) Full {B}ayesian inference with hazard mixture
  models. To appear in Comput Stat Data An
  \urlprefix\url{http://dx.doi.org/10.1016/j.csda.2014.12.003}

\bibitem[{Doksum(1974)}]{Doksum}
Doksum K (1974) Tailfree and neutral random probabilities and their posterior
  distributions. Ann Probab 2(2):183--201

\bibitem[{Dykstra and Laud(1981)}]{DykLau81}
Dykstra R, Laud P (1981) {A {B}ayesian nonparametric approach to reliability}.
  Ann Statist 9(2):356--367

\bibitem[{Ferguson(1973)}]{ferguson1973bayesian}
Ferguson T (1973) {A Bayesian analysis of some nonparametric problems}. Ann
  Statist 1(2):209--230

\bibitem[{Gelfand and Kottas(2002)}]{GelKot02}
Gelfand AE, Kottas A (2002) {A computational approach for full nonparametric
  {B}ayesian inference under Dirichlet process mixture models}. J Comput Graph
  Stat 11(2):289--305

\bibitem[{Hjort(1990)}]{Cervo}
Hjort N (1990) Nonparametric {B}ayes estimators based on beta processes in
  models for life history data. Ann Statist 18(3):1259--1294

\bibitem[{Ishwaran and James(2004)}]{IshJam04}
Ishwaran H, James L (2004) Computational methods for multiplicative intensity
  models using weighted gamma processes: proportional hazards, marked point
  processes, and panel count data. J Am Stat Assoc 99(465):175--190

\bibitem[{James(2005)}]{Jam05}
James L (2005) {Bayesian Poisson process partition calculus with an application
  to Bayesian L\'evy moving averages}. Ann Statist 33(4):1771--1799

\bibitem[{Jara et~al(2011)Jara, Hanson, Quintana, M{\"u}ller, and
  Rosner}]{jara2011dppackage}
Jara A, Hanson T, Quintana F, M{\"u}ller P, Rosner G (2011) {DPpackage:
  Bayesian non-and semi-parametric modelling in R}. J Stat Softw 40(5):1

\bibitem[{Lo and Weng(1989)}]{LoWen89}
Lo A, Weng C (1989) {On a class of Bayesian nonparametric estimates. II. Hazard
  rate estimates}. Ann I Stat Math 41(2):227--245

\bibitem[{Provost(2005)}]{provost2005moment}
Provost SB (2005) Moment-based density approximants. Mathematica J
  9(4):727--756

\end{thebibliography}

\end{document}